\documentclass[twocolumn,eqsecnum,nobibnotes,amsmath,amssymb,aps,prb,showpacs,floats,reprint]{revtex4-1}
\usepackage{graphicx}

\begin{document}
\title{Quasiclassical description of a superconductor with a spin density
wave}
\author{A. Moor, A. F. Volkov, and K. B. Efetov}
\affiliation{Theoretische Physik III,\\
Ruhr-Universit\"{a}t Bochum, D-44780 Bochum, Germany\\}

\begin{abstract}
We derive equations for the quasiclassical Green's functions $\check{g}$ within
a simple model of a two-band superconductor with a spin-density-wave (SDW).
The elements of the matrix $\check{g}$ are the retarded, advanced, and Keldysh functions
each of which is an $8\times 8$ matrix in the Gor'kov-Nambu,
the spin and the band space. In equilibrium, these equations are a
generalization of the Eilenberger equation. On the basis of the derived
equations we analyze the Knight shift, the proximity and the dc Josephson
effects in the superconductors under consideration. The Knight shift is
shown to depend on the orientation of the external magnetic field with respect to
the direction of the vector of the magnetization of the SDW. The proximity effect
is analyzed for an interface between a superconductor with the SDW and a
normal metal. The function describing both superconducting and magnetic
correlations is shown to penetrate the normal metal or a metal with the SDW
due to the proximity effect. The dc Josephson current in an $S_{SDW}/N/S_{SDW}$
junction is also calculated as a function of the phase difference $\varphi$.
It is shown that in our model the Josephson current does not depend on the
mutual orientation of the magnetic moments in the superconductors $S_{SDW}$
and is proportional to~$ \sin \varphi$. The dissipationless spin current $j_{sp}$
depends on the angle $\alpha $ between the magnetization vectors in the
same way ($j_{sp} \sim \sin \alpha $) and is not zero above the
superconducting transition temperature.
\end{abstract}

\date{\today}
\pacs{74.20 Fg, 74.20 Rp, 74.72.-h, 75.40 Gb, 74.25 Dw, 74.25 Ha}
\maketitle

\section{Introduction}

Coexistence of two or more order parameters in solids is one of the most
intriguing phenomena in condensed matter physics. There are many systems
where the ordered phases are even antagonistic to each other. Nowadays,
a very popular subject of research are compounds with coexisting
superconducting (SC) and magnetic (M) order parameters.

It is already well known, that the exchange field responsible for
ferromagnetism destroys singlet Cooper pairs \cite{BuzdinRMP} but can
generate \cite{BVErmp,Eschrig} triplet ones (odd in frequency and symmetric in
space) that survives even in the presence of a strong exchange field
and strong impurity scattering. The theoretical prediction has been recently
confirmed experimentally \cite{Klapwijk,Sosnin,Birge,Robinson,Westerholt,Aarts}.

In contrast, a magnetic order of the antiferromagnetic type can
coexist with singlet Cooper pairs because the average magnetic moment
(on distances of order of the size of the Cooper pairs) is close to
zero. This case is realized in superconductors with a spiral magnetic
structure (for a review, see \cite{BulaevRev} and references therein).

Another example of this coexistence are superconductors with a
spin-density wave (SDW) \cite{Tugushev,Kopaev,Littlewood}. This type of
superconductivity coexisting with the SDW is realized in
quasi-one-dimensional conductors (for a review, see \cite{Chaikin}).
A lot of attention is payed now to quasi-two-dimensional superconducting
compounds -- the so-called Fe-based pnictides -- discovered recently
\cite{Discov1,Discov2,Discov3,Discov4}. Some of these Fe-based superconductors
have rather high critical temperature $T_{c}$ of the superconducting
transition (up to 56 K observed in $Gd_{1-x}Th_{x}FeAsO$ at the doping level
$x=0.2$ \cite{Wang}) and have much in common with the high $T_{c}$ cuprates
(see, for example, \cite{Review1,Review2,Kotliar,Sachdev} and references
therein). It has been established, both theoretically and
experimentally, that at a certain critical temperature $T_{M}$ (usually
$T_{M}>T_{c}$) the compounds undergo a magnetic transition leading to the
formation of a SDW. In a certain interval of the temperature and
doping level $x$ the SC and M order parameters may coexist
\cite{Mazin08,Chubukov09,Chubukov10,Schmalian10,Review1}.

The SDW in pnictides arises due to the nesting of the electron and hole
pockets \cite{Tesanovic,Schmalian10,Chubukov09,Review1,Review2}. Each pocket
has its own SC order parameter, $\Delta_{el}$ resp.\ $\Delta _{h}$, but the most
energetically favorable state corresponds to the so-called $s_{{+-}}$-pairing,
characterized by the opposite signs of the SC order parameters:
$\Delta_{el}=-\Delta_{h}$ \cite{Mazin08,Chubukov09,Schmalian10,Scalapino,Choi}.

Microscopic theories for the electronic states in materials with the
superconducting and antiferromagnetic order parameters are mainly based on
the equations for the Green's functions $G$ which include both normal and
anomalous ones corresponding to the SC and M order parameters
\cite{Chubukov09,Chubukov10,Schmalian10,Eremin10}. These equations are written
in the mean field approximation in an analogy with the ordinary superconductors
without specifying the microscopic mechanism of superconductivity and were
solved under some approximations mostly in spatially homogenous case.
On the other hand, several research groups have begun to study pnictides in
nonhomogeneous structures. For examples, the results concerning Josephson effects
in junctions based on pnictides have been published recently
\cite{Joseph1,Joseph2,Joseph3,Joseph4}. The Josephson effect in tunnel $S/I/S$
junctions ($I$ denotes an insulator) consisting of the of multiband
superconductors can be investigated on the basis of the tunnel Hamiltonian
and the equations for the Green's functions $G$ \cite{Parker,Chen,Sudbo09}. However,
this approach is not applicable to other types of the Josephson junctions.

In order to tackle the problems in nonhomogeneous cases it is much more
convenient to employ the widely used quasiclassical approach based on using
the so-called quasiclassical Green's functions $g$. These functions are obtained
from the full Green's functions $G$ by integration over the modulus of the momentum
$\mathbf{p}$ in the vicinity of the Fermi surface.

The quasiclassical Green's functions $g$ are used in situations when the parameters
characterizing the system vary on the distances exceeding the Fermi wave length
$\lambda_{F}$. The method of quasiclassical Green's functions has been developed in
the theory of superconductivity \cite{Eilenberger,Usadel,LO} and turned out to be
the most powerful and effective tool in dealing with the problems in nonuniform cases:
vortices in superconductors, proximity effect in superconductor/normal metal
structures, etc. The equations for the quasiclassical Green's functions can
be generalized to the case of the charge-density-wave (CDW)
\cite{ArtVolkovCDW,Gor'kov} and two-band superconductors
\cite{Koshelev03,Vorontsov07,Anishchanka07,Gurevich10}. They are very efficient
for describing superconductors in contact with ferromagnets \cite{BVErmp}.

At the same time, corresponding equations for the case of a two-band
superconductor with the SDW are still lacking, although such equations can
be derived under certain restrictions. Deriving these equations is very
important because they can serve for the description of the superconductors like
Fe-based pnictides. The quasiclassical approximation is well justified for
describing the pnictides because the correlation lengths for both the
superconducting and the magnetic correlations $\xi_{S,M}~\approx~v\hbar/\Delta,v\hbar/W_{M}$
are much longer than the Fermi wave length, where $W_{M}~\sim~T_{M}$
is a characteristic energy related to the magnetic order.

In this paper, we derive the equations for the quasiclassical matrix Green's
function $\check{g}$ that describe a two-band superconductor with the SDW.
These equations can be applied both to equilibrium and nonequilibrium states
in homogeneous and nonhomogeneous cases and describe a broad class of
phenomena in superconductors like Fe-based pnictides. We employ these
equations to analyse the Knight shift, the proximity and the dc Josephson
effect in such superconductors.

The paper is organized as follows. In Sec.~II we write the Hamiltonian of
the system in the mean-field approximation in terms of the operators suitable
for deriving the equations for the quasiclassical Green's functions. These
equations are derived in Sec.~III in the ballistic limit. We also present
formulas for the observable quantities such as the SC and M order parameters
($\Delta$ and $W_{M}$), as well as for the charge and spin current
density. In Sec.~IV we study the Knight shift, i.e.\ the shift of the NMR
(nuclear magnetic resonance) frequency due to the spin polarization of the
s-electrons, for different orientation of the external magnetic field with
respect to the direction of the magnetization $\mathbf{m}$ in the SDW. In Sec.~V the
proximity effect will be analyzed in the vicinity of the interface between a
superconductor with the SDW and a nonsuperconducting material (with or without
SDW). Using a simple model, in Sec.~VI we calculate the dc Josephson ($\mathbf{j}_{J}$)
and the spin ($\mathbf{j}_{sp}$) current in the Josephson $S_{SDW}/N/S_{SDW}$ junction,
where $S_{SDW}$ denotes a superconductor with an SDW, and $N$ is a normal
metal. The obtained results will be discussed in Sec.~VII.

\section{Model and Basic Equations}

We consider a simple model of a two band superconductor with such a Fermi
surface that not only the superconducting but also an SDW pairing is possible.
The SDW pairing may originate from the nesting of certain parts of the Fermi
surface and we assume that such parts exist.

In such a situation one can have logarithmic contribution not only
from the Cooper channel but also from the particle-hole one. Solving this
problem microscopically is not easy because one should perform rather
complicated renormalization group calculations in order to get the
information about non-trivial phases at low temperature.

We do not intend to discuss here microscopic mechanisms of the
superconductivity and its competition with the SDW. Our goal is more modest: assuming
that the superconductivity and the SDW coexist we consider them in the
mean field approximation. In principle, one could obtain physical quantities
of interest using the normal and anomalous electron Green's functions written
in the presence of the superconducting and SDW order parameters \cite{Chubukov09,Schmalian10}.

Unfortunately, if the system is not homogeneous, solving the equations
for the electron Green's functions is not easy and therefore we develop
the formalism of quasiclassical Green's functions $g$ taking into account
both the superconducting and the SDW order parameters. These equations are
obtained as a result of a certain simplification of the original equations
for the electron Green's functions. This approach is valid in the situations
when physical quantities vary slowly on the distances of the order of the
electron wavelength.

To be close to the experimental results on pnictides we assume that the
$s$-wave superconducting pairing inside the bands is most important. The
spins of the electrons interact with both the exchange field of the SDW and
the external field $\mathbf{H}$.

The Hamiltonian $\mathcal{H}$ of the system under consideration can be
written in the form (see Refs.~\cite{Chubukov09,Schmalian10})
\begin{widetext}
\begin{align}
\mathcal{H} = \sum_{\mathbf{p}, \alpha, \beta} & \Bigl\lbrace
        \xi_{1}(\mathbf{p}) \hat{c}_{1\alpha}^{\dagger} \hat{c}_{1\alpha} +
        \xi_{2}(\mathbf{p}) \hat{c}_{2\alpha}^{\dagger} \hat{c}_{2\alpha} +
        \frac{1}{2}(i\hat{\sigma}_{2})_{\alpha \beta}
        \bigl( \Delta_{1} \hat{c}_{1\alpha}^{\dagger} \hat{c}_{1\beta}^{\dagger} +
        \Delta_{2} \hat{c}_{2\alpha}^{\dagger} \hat{c}_{2\beta}^{\dagger} - h.c. \bigr) +
        W_{M0} (\hat{\sigma}_{3})_{\alpha \beta} \bigl( \hat{c}_{1\alpha}^{\dagger} \hat{c}_{2\beta} +
        \hat{c}_{2\alpha}^{\dagger} \hat{c}_{1\beta} \bigr)
\label{Ham1} \\
    &-W_{Z1} \bigl[ (\hat{\sigma}_{3})_{\alpha \beta} \cos\theta + (\hat{\sigma}_{1})_{\alpha \beta}
        \sin\theta \bigr] \hat{c}_{1\alpha}^{\dagger} \hat{c}_{1\beta} -
        W_{Z2} \bigl[ (\hat{\sigma}_{3})_{\alpha \beta} \cos\theta +
        (\hat{\sigma}_{1})_{\alpha \beta} \sin\theta \bigr] \hat{c}_{2\alpha}^{\dagger}
        \hat{c}_{2\beta} \Bigr\rbrace \notag \,,
\end{align}
\end{widetext}
where $\xi _{1,2}(\mathbf{p})$ is the kinetic energy in the $1$ (hole) and
$2$ (electron) bands, respectively, counted from the Fermi energy and
$\Delta_{1,2}$ is the superconducting order parameter in these bands.

The quantity $W_{M0}=(W_{M}+W_{M}^{\ast})/2$ is the magnetic order
parameter describing the SDW (for simplicity we set the incommensurability
wave vector $\mathbf{q}$ to zero), $W_{M}~=~\mu _{eff}m$, where $\mu_{eff}$
is the effective magnetic moment of the free electrons participating in the
formation of the SDW and $m$ is the magnetization of the SDW.

The terms $W_{Z1}=\mu_{1}H$ and $W_{Z2}=\mu_{2}H$ are the Zeeman energies
in the presence of the external magnetic field with the components
$\mathbf{H}~=~H(\sin\theta ,0,\cos\theta)$, where $\mu_{1}$, $\mu_{2}$ are the
effective magnetic moments in the hole and electron bands, respectively. We
introduce the Pauli matrices $\hat{\rho}_{i}$, $\hat{\tau}_{i}$, $\hat{\sigma}_{i}$
operating in the ``band'', Gor'kov-Nambu and spin spaces, respectively;
$\hat{\rho}_{0}$, $\hat{\tau}_{0}$, $\hat{\sigma}_{0}$ being the corresponding
unit matrices.

We assume that the functions $\xi_{1,2}(\mathbf{p})$ have the form
$\xi_{1,2}(\mathbf{p})~=~\mp \xi(\mathbf{p})+\delta \mu$, where the parameter
$\delta \mu$ describes the deviation from the ideal nesting depending on doping.
Thus, the band 1 and 2 are the hole and electron bands, respectively.
The energy $\xi(\mathbf{p})$ can be linearized near the Fermi surface and we write
it in the form $\xi(\mathbf{p})~=~v_{F}(|\mathbf{p|}-p_{F})$ with $v_{F}=p_{F}/m$
assuming for simplicity that the Fermi velocities $v_{F}~\approx~|\mathbf{v}|$
in the bands are equal to each other. Actually, this assumption is not fulfilled
in real pnictides but the results obtained under this assumption are valid, at
least qualitatively, even in the case of unequal Fermi velocities.

For convenience we introduce new operators, $\hat{a}_{\alpha}$ and $\hat{b}_{\alpha}$,
related to the operators $\hat{c}_{n\alpha}^{\dagger}, \hat{c}_{n\alpha}$

\begin{equation}
\hat{a}_{\bar{\alpha}}=\hat{c}_{1\alpha}^{\dagger}\,, \qquad
\hat{b}_{\alpha}=\hat{c}_{2\alpha} \,,
\label{a,b-operator}
\end{equation}
where

\begin{equation}
\bar{\alpha} = \begin{cases}
                2 \,, & \text{if } \alpha=1 \\
                1 \,, & \text{if } \alpha=2
                \end{cases} \,.
\end{equation}

Then, one can write the Hamiltonian, Eq.~(\ref{Ham1}), as follows.

The kinetic energy part $\mathcal{H}_{kin}$ reads as

\begin{align}
\mathcal{H}_{kin} = &\sum_{\mathbf{p},\alpha\beta} \Bigl\lbrace
(\hat{\sigma}_0)_{\alpha\beta} \Bigl[ \xi(\mathbf{p})
\bigl(\hat{a}_{\alpha}^{\dagger}\hat{a}_{\beta} + \hat{b}_{\alpha}^{\dagger}
\hat{b}_{\beta} \bigr)
\label{E kin} \\
&- \delta\mu \bigl(\hat{a}_{\alpha}^{\dagger}\hat{a}_{\beta} -
\hat{b}_{\alpha}^{\dagger}\hat{b}_{\beta} \bigr) \Bigr] + h.c. \Bigr\rbrace \,. \notag
\end{align}

The term describing the superconducting pairing takes the form

\begin{equation}
\mathcal{H}_{SC} = -\frac{1}{2} \sum_{\mathbf{p},\alpha,\beta} \left\lbrace
i (\hat{\sigma}_2)_{\alpha\beta} \bigl(\Delta_{a} \hat{a}_{\alpha}^{\dagger}
\hat{a}_{\beta}^{\dagger} + \Delta_{b} \hat{b}_{\alpha}^{\dagger}
\hat{b}_{\beta}^{\dagger} - h.c. \bigr) \right\rbrace \,,
\label{E_super}
\end{equation}
where $\Delta_{a}=\Delta_{1}^{\ast}$ and $\Delta_{b}=\Delta_{2}$.

The term related to the SDW can be written as follows

\begin{equation}
\mathcal{H}_{SDW} = \sum_{\mathbf{p},\alpha,\beta} \left\lbrace i
(\hat{\sigma}_2)_{\alpha\beta} W_{M0} \hat{a}_{\alpha}^{\dagger}
\hat{b}_{\beta}^{\dagger} + h.c. \right\rbrace \,.
\label{E_SDW}
\end{equation}

At last, the Zeeman term can be written as

\begin{widetext}
\begin{align}
\mathcal{H}_{Z} &= -\sum_{\mathbf{p},\alpha ,\beta } \left\{ W_{Z1} \bigl(
(\hat{\sigma}_{3})_{\alpha \beta} \cos\theta -(\hat{\sigma}_{1})_{\alpha \beta}
\sin \theta \bigr) \hat{a}_{\alpha}^{\dagger}\hat{a}_{\beta}
+ W_{Z2} \bigl( (\hat{\sigma}_{3})_{\alpha \beta} \cos\theta +
(\hat{\sigma}_{1})_{\alpha \beta} \sin\theta \bigr)
\hat{b}_{\alpha}^{\dagger} \hat{b}_{\beta }\right\}
\label{E_Z} \\
&= -\sum_{\mathbf{p},\alpha ,\beta} W_{Z} \left\{ \bigl(
(\hat{\sigma}_{3})_{\alpha \beta} \cos \theta -(\hat{\sigma}_{1})_{\alpha \beta}
\sin\theta \bigr) \hat{a}_{\alpha}^{\dagger} \hat{a}_{\beta } +
\bigl( (\hat{\sigma}_{3})_{\alpha \beta} \cos\theta + (\hat{\sigma}_{1})_{\alpha \beta}
\sin\theta \bigr) \hat{b}_{\alpha}^{\dagger} \hat{b}_{\beta} \right\} \,, \notag
\end{align}
\end{widetext}
where we set $W_{Z1}=W_{Z2}=W_{Z}=\mu_{eff}H$.

Eqs.~(\ref{E kin}--\ref{E_Z}) are cumbersome and it is not convenient to use
them directly. Fortunately, they can be rewritten in a more compact form
introducing the operators $\hat{A}_{n\alpha }$ and $\hat{B}_{n\alpha }$
which are matrices in Gor'kov-Nambu (index $n$) and spin (index $\alpha $) space
(see for example \cite{BVErmp}):

\begin{align}
\label{A,B-operator}
\hat{A}_{1\alpha} &= \hat{a}_{\alpha}\,, & \hat{A}_{2\bar{\alpha}}
    &= \hat{a}_{\alpha}^{\dagger} \,; \\
\hat{B}_{1\alpha} &= \hat{b}_{\alpha}\,, & \hat{B}_{2\bar{\alpha}}
&= \hat{b}_{\alpha}^{\dagger} \,. \notag
\end{align}

In order to take into account two bands or, in other words, different parts
of the Fermi surface, we define operators $\hat{C}_{mn\alpha}$ with the index
$m$ related to different bands so that

\begin{equation}
\hat{C}_{1n\alpha}=\hat{A}_{n\alpha}\,, \quad \hat{C}_{2n\alpha}=\hat{B}_{n\alpha} \,.
\label{C-operator}
\end{equation}

The operators $\hat{C}_{mn\alpha}$ obey the commutation relations

\begin{align}
\hat{C}_{km\alpha}^{\dagger}\hat{C}_{ln\beta}+\hat{C}_{ln\beta}\hat{C}_{km\alpha}^{\dagger}
    &= \delta_{kl}\delta_{mn}\delta_{\alpha\beta} \,, \\
\hat{C}_{km\alpha}\hat{C}_{ln\beta}+\hat{C}_{ln\beta}\hat{C}_{km\alpha}
    &= \delta_{kl}\delta_{m\bar{n}}\delta_{\alpha\bar{\beta}} \,.
\label{C-commut}
\end{align}

After rewriting the energy terms Eqs.~(\ref{E kin}--\ref{E_Z}) in terms of
the operators $\{\hat{C}_{mn\alpha }^{\dagger },\hat{C}_{mn\alpha }\}~=~\{\hat{C}^{\dagger },\hat{C}\}$
the Hamiltonian in Eq.~(\ref{Ham1}) can be written in the following way

\begin{equation}
\mathcal{H} = \frac{1}{2} \sum_{\mathbf{p}}\hat{C}^{\dagger}\hat{\mathrm{H}}\hat{C} \,,
\label{Ham2}
\end{equation}
where the operator $\hat{\mathrm{H}}$ has the form

\begin{equation}
\label{H-operator}
\hat{\mathrm{H}} = \hat{\mathrm{H}}_{kin} + \hat{\mathrm{H}}_{SC} +
\hat{\mathrm{H}}_{SDW} + \hat{\mathrm{H}}_{Z}
\end{equation}
with

\begin{align}
&\hat{\mathrm{H}}_{kin} =  \xi(\mathbf{p}) \cdot
    \hat{\rho}_{0} \cdot \hat{\tau}_{3} \cdot \hat{\sigma}_{0} - \delta\mu \cdot
    \hat{\rho}_{3} \cdot \hat{\tau}_{3} \cdot \hat{\sigma}_{0}  \,, \\
&\hat{\mathrm{H}}_{SC} =
    \begin{cases}
        \Delta^{\prime} \hat{\rho}_{0} \cdot \hat{\tau}_{1} \cdot \hat{\sigma}_{3} +
        i \Delta^{\prime \prime} \hat{\rho}_{3} \cdot i \hat{\tau}_{2} \cdot
        \hat{\sigma}_{3} \,, & s_{++} \\
        \Delta^{\prime} \hat{\rho}_{3} \cdot \hat{\tau}_{1} \cdot \hat{\sigma}_{3} +
        i \Delta^{\prime \prime} \hat{\rho}_{0} \cdot i \hat{\tau}_{2} \cdot
        \hat{\sigma}_{3} \,, & s_{+-} \\
    \end{cases} \,, \\
&\hat{\mathrm{H}}_{SDW(z)} = W_{M0} \cdot
    \hat{\rho}_{1} \cdot \hat{\tau}_{1} \cdot \hat{\sigma}_{3}\,, \\
&\hat{\mathrm{H}}_{Z} = -W_{Z} \cdot \{\hat{\rho}_{0}
    \cdot \hat{\tau}_{0} \cdot \hat{\sigma}_{3}\cos\theta - \hat{\rho}_{3}\cdot
    \hat{\tau}_{3} \cdot \hat{\sigma}_{1} \sin\theta \} \,.
\label{Hkin-HZ}
\end{align}

The order parameter $\Delta \equiv \Delta^{\prime }+i\Delta^{\prime \prime}$ is
related to $\Delta_{1,2}$ as $\Delta_{1}=\Delta^{\ast}=\Delta_{2}$ ($s_{++}$-pairing)
and $\Delta_{1}=\Delta^{\ast}=-\Delta _{2}$ ($s_{+-}$-pairing).

One can perform a rotation in the spin space and change the direction of the
field $\mathbf{H}_{ex}$ and the magnetization $\mathbf{m}$. The rotation around
the axis $j=x,y,z$ by the angle $\vartheta $ means a unitary transformation
of the Hamiltonian $\hat{\mathrm{H}}$,

\begin{equation}
\hat{\mathrm{H}} \Rightarrow \hat{R}_{j}\hat{\mathrm{H}}
\hat{R}_{j}^{\dagger}\,,
\label{RotTrans}
\end{equation}
where the unitary rotation matrix $\hat{R}_{j}$ has the form
\begin{equation}
\hat{R}_{j}=\cos(\vartheta /2)+\hat{r}_{j}\sin(\vartheta /2)
\end{equation}
and

\begin{equation}
\hat{r}_{j} =
    \begin{cases}
    i \cdot \hat{\rho}_{3} \cdot \hat{\tau}_{3} \cdot \hat{\sigma}_{1} \,, & j=x
    \\
    i \cdot \hat{\rho}_{3} \cdot \hat{\tau}_{3} \cdot \hat{\sigma}_{2} \,, & j=y
    \\
    i \cdot \hat{\rho}_{0} \cdot \hat{\tau}_{0} \cdot \hat{\sigma}_{3} \,, & j=z
    \end{cases} \,.
\end{equation}

If the magnetization in the SDW is oriented in the $x$- or $y$-direction,
then $\hat{\mathrm{H}}_{SDW}$ has the form

\begin{equation}
\hat{\mathrm{H}}_{SDW(x,y)} = \pm W_{M0} \cdot \hat{\rho}_{2}
\cdot \hat{\tau}_{2} \cdot \hat{\sigma}_{1,2} \,.
\label{HSDWx,y}
\end{equation}

Having specified the Hamiltonian of the model in the compact matrix notation
we can introduce the Green's functions in terms of the operators
$\hat{C}^{\dagger },\hat{C}$ in the usual way. For example, the retarded
Green's function $\hat{G}^{R}$ can be written as

\begin{widetext}
\begin{align}
\hat{G}^{R}(\mathbf{p},\mathbf{p}^{\prime };t,t^{\prime }) &= (1/i) \left\langle
\check{C}(\mathbf{p};t)\cdot \check{C}^{\dagger }(\mathbf{p}^{\prime};t^{\prime })
+\check{C}^{\dagger }(\mathbf{p}^{\prime };t^{\prime })\cdot
\check{C}(\mathbf{p};t) \right\rangle \theta (t-t^{\prime }) \,,
\label{G^R}
\\
\intertext{while the Keldysh function $\hat{G}^{K}$ takes the form}
\hat{G}^{K}(\mathbf{p},\mathbf{p}^{\prime };t,t^{\prime }) &= (1/i) \left\langle
\check{C}(\mathbf{p};t)\cdot \check{C}^{\dagger}(\mathbf{p}^{\prime};t^{\prime})
-\check{C}^{\dagger}(\mathbf{p}^{\prime};t^{\prime})
\cdot \check{C}(\mathbf{p};t) \right\rangle \,.
\label{G^K}
\end{align}
\end{widetext}

One can define the matrix Green's function $\check{G}$ with the block elements
$\hat{G}^{R}$, $\hat{G}^{K}$ and $\hat{G}^{A}$ (see \cite{LO,BelzigRev,Rammer,Kopnin,BVErmp})

\begin{equation}
\label{SuperG}
\check{G}=
    \begin{pmatrix}
    \hat{G}^{R} & \hat{G}^{K} \\
    0 & \hat{G}^{A}
    \end{pmatrix} \,.
\end{equation}

Using these functions we can calculate various macroscopic quantities. The Green's
functions obey dynamic equations analogous to the Eilenberger equation and their
generalizations to the nonequilibrium case. \cite{LO,BelzigRev,Kopnin,Rammer,BVErmp}

\section{Equations for quasiclassical Green's functions}

Following the method of the quasiclassical Green functions
\cite{LO,BelzigRev,Rammer,Kopnin,BVErmp} one should introduce the quasiclassical
matrix Green's function $\check{g}$ related to the Green's function
$\check{G}(r,r;t,t^{\prime})$ as

\begin{equation}
\check{g}=(i/\pi )\int d\xi \, \left(\hat{\tau}_{3} \cdot \check{G} \right)
\label{Quasicl}
\end{equation}
and derive dynamic equations for it. This section is devoted to such a
derivation that can be carried out in the standard way.

The equations for the matrix $\check{G}$ can be obtained from the equation
of motion for the operators $\hat{C}(\mathbf{k};t)$,

\begin{equation}
i\partial_{t} \hat{C}(\mathbf{p};t) = \bigl\lbrack \hat{C}(\mathbf{p};t)\,,\mathcal{H} \bigr\rbrack\ \,,
\label{Heisenberg}
\end{equation}
and the definition of the Green's functions $\hat{G}^{R}$, $\hat{G}^{K}$ and
$\hat{G}^{A}$, Eqs.~(\ref{G^R}--\ref{G^K}).

Taking into account the commutation relations (\ref{C-commut}), we obtain

\begin{equation}
i \partial_{t} \check{G} - \hat{\mathrm{H}} \cdot \check{G} =
\check{1} \delta(t-t^{\prime}) \,,
\label{EqforG}
\end{equation}
where the matrix $\mathrm{\hat{H}}$ is defined in Eqs.~(\ref{H-operator}--\ref{Hkin-HZ}).

When deriving Eq.~(\ref{EqforG}), we used the property
$\bigl( \hat{\mathrm{H}}\bigr)_{ms\alpha}^{nt\beta}~=~-\bigl(\hat{\mathrm{H}}\bigr)_{n\bar{t}\bar{\beta}}^{m\bar{s}\bar{\alpha}}$.

Analogously, one can obtain the conjugate equation

\begin{equation}
-i \partial_{t^{\prime}} \check{G} - \check{G} \cdot
\hat{\mathrm{H}} = \check{1}\delta (t-t^{\prime}) \,.
\label{EqforGcc}
\end{equation}

The next steps are standard for deriving the Eilenberger equation \cite{Eilenberger}
or the equation for a more general Green's function $\check{G}$
\cite{LO,Rammer,BelzigRev,Kopnin,BVErmp}. We multiply Eq.~(\ref{EqforG}) by
$\hat{\tau}_{3}$ from the left and Eq.~(\ref{EqforGcc}) from the right and
subtract these equations from each other. Then, the obtained equation is
integrated over the energy $\xi(p)$ where $p~=~|(\mathbf{p}+\mathbf{p}^{\prime })|/2$.
Finally, we obtain the equation for the matrix Green's function

\begin{equation}
\bigl(\hat{\tau}_{3} \cdot \partial_{t} \check{g} +
\partial_{t^{\prime}} \check{g} \cdot \hat{\tau}_{3} \bigr) +
\mathbf{v} \nabla \check{g} + i \bigl\lbrack \hat{\mathrm{\Lambda}}^{(j)}_{\pm} \,,
\check{g} \bigr\rbrack = 0 \,,
\label{Eilenberger}
\end{equation}
where

\begin{equation}
\hat{\mathrm{\Lambda}}^{(j)}_{\pm} = \hat{\mathrm{h}}_{\mu} + \hat{\mathrm{h}}_{SC}^{\pm} +
\hat{\mathrm{h}}_{SDW}^{(j)} - \hat{\mathrm{h}}_{Z}
\label{Lambda}
\end{equation}
with (``$\pm$'' stands for $s_{++}$-, $s_{+-}$-pairing)

\begin{align}
&\hat{\mathrm{h}}_{\mu} = \delta\mu \cdot \hat{\rho}_{3} \cdot
    \hat{\tau}_{0} \cdot \hat{\sigma}_{0} \,, \\
&\hat{\mathrm{h}}_{SC}^{\pm} = - \Delta^{\prime} \cdot \hat{\rho}_{0,3}
    \cdot i \hat{\tau}_{2} \cdot \hat{\sigma}_{3} -
    i \Delta^{\prime \prime} \cdot \hat{\rho}_{3,0} \cdot \hat{\tau}_{1} \cdot \hat{\sigma}_{3} \,, \\
&\hat{\mathrm{h}}_{SDW}^{(j)} =
    \begin{cases}
        i \cdot W_{M0} \cdot \hat{\rho}_{2} \cdot \hat{\tau}_{1}
            \cdot \hat{\sigma}_{1} \,, & j=x \\
        - i \cdot W_{M0} \cdot \hat{\rho}_{2} \cdot \hat{\tau}_{1}
            \cdot \hat{\sigma}_{2} \,, & j=y \\
        i \cdot W_{M0} \cdot \hat{\rho}_{1} \cdot \hat{\tau}_{2}
            \cdot \hat{\sigma}_{3} \,, & j=z
    \end{cases} \,, \\
&\hat{\mathrm{h}}_{Z} = W_{Z} \cdot \hat{\mathit{P}} \,,
\label{DefnLambda}
\end{align}
where the matrix $\hat{\mathit{P}}=\hat{\mathit{P}}_{1}+\hat{\mathit{P}}_{3}$
with $\hat{\mathit{P}}_{1}=\hat{\rho}_{3}\cdot \hat{\tau}_{0}\cdot
\hat{\sigma}_{1}\sin \theta $ and $\hat{\mathit{P}}_{3}=-\hat{\rho}_{0}
\cdot \hat{\tau}_{3}\cdot \hat{\sigma}_{3}\cos \theta$ describes the
external magnetic field.

Using the Matsubara representation one reduces Eq.~(\ref{Eilenberger}) to the form

\begin{equation}
\bigl\lbrack \omega_{n} \hat{\tau}_{3}\,,\check{g} \bigr\rbrack + \mathbf{v} \nabla \check{g}+
i \bigl\lbrack \hat{\mathrm{\Lambda}}^{(j)}_{\pm} \,, \check{g} \bigr\rbrack = 0 \,.
\label{EilenbergerMatsubara}
\end{equation}

Repeating arguments used in the derivation for superconductors
\cite{LO,Rammer,BelzigRev,Kopnin,BVErmp} or two-band metals \cite{ArtVolkovCDW}
a normalization condition for the matrix $\check{g}$ can easily be derived

\begin{equation}
\check{g}(t,t_{1}) \cdot \check{g}(t_{1},t^{\prime }) = \check{1} \delta
(t-t^{\prime })
\label{NormCond}
\end{equation}
with $\bigl(\check{1}\bigr)_{ms\alpha}^{nt\beta} = \delta_{mn} \delta_{st}
\delta_{\alpha\beta}$.

Eqs.~(\ref{Eilenberger}, \ref{NormCond}) supplemented by proper boundary
conditions allow one to find unambiguously the quasiclassical Green's function
$\check{g}$. As soon as this matrix Green's function is known, one can calculate
macroscopic quantities of interest. For example the current density
$\mathbf{j}_{\omega}$ in the system is equal to

\begin{equation}
\mathbf{j}_{\omega}=\frac{1}{8}e\nu \int d\epsilon \, \mathrm{Tr}
\bigl(\hat{\rho}_{3}\cdot \hat{\tau}_{3}\cdot \hat{\sigma}_{0}
\cdot \bigl\langle\mathbf{v}\check{g}(\epsilon ,\epsilon^{\prime })\bigr\rangle\bigr)\,,
\label{Current}
\end{equation}
where $\epsilon=(\epsilon+\epsilon^{\prime })/2$, $\omega=\epsilon-\epsilon^{\prime }$,
$\nu =p_{F}m/\pi^{2}$ is the density-of-states at the Fermi energy and the angle brackets
mean the averaging over the momentum directions:
\begin{equation}
    \bigl\langle\bigl(\dots\bigr)\bigr\rangle = \int \frac{d \Omega}{4 \pi} \, \bigl(\dots\bigr) \,.
\end{equation}

The components of the static magnetic moment $\mathbf{M}$ in
the $x$-, $y$- and $z$-directions are given by

\begin{align}
M_{x,y} &= M_{0}-i(2\pi T)\nu \mu _{B}\frac{1}{8}\sum_{\omega =0}^{\infty}
\mathrm{Tr}\bigl(\hat{\rho}_{3}\cdot \hat{\tau}_{0}\cdot \hat{\sigma}_{1,2}\cdot
\check{g}(\omega )\bigl) \,,
\label{Mx,y} \\
M_{z} &= M_{0}-i(2\pi T)\nu \mu _{B}\frac{1}{8}\sum_{\omega =0}^{\infty }
\mathrm{Tr}\bigl(\hat{\rho}_{0}\cdot \hat{\tau}_{3}\cdot \hat{\sigma}_{3}\cdot
\check{g}(\omega )\bigr) \,,
\label{Mz}
\end{align}%
where $M_{0}=\mu_{eff}^{2}\nu H$ is the Pauli paramagnetic term. This
term arises as a result of the integration over momenta far from the Fermi
surface \cite{BVErmp} and cannot be calculated in the quasiclassical
approximation. Writing Eqs.~(\ref{Mx,y}, \ref{Mz}) we replaced the integration
over the energy~$\epsilon $ by the summation over the Matsubara frequencies.
Below we are interested in the polarization of electron spins by a static
magnetic field $\mathbf{H}$ and this is why Eqs.~(\ref{Mx,y}, \ref{Mz})
are written for the static case.

Using Eq.~(\ref{Eilenberger}) one can obtain the expressions for the spin
currents $\mathbf{j}_{sp}$. For example, the spin currents with $(x,y)$%
-spin projections are given by

\begin{equation}
\mathbf{j}_{sp}^{(x,y)}=-i(2\pi T)\frac{1}{8}\mu_{B} \nu
\sum_{\omega=0}^{\infty} \mathrm{Tr}\bigl(\hat{\rho}_{3} \cdot \hat{\tau}_{3} \cdot \hat{\sigma}_{1,2}
\cdot \bigl\langle \mathbf{v} \check{g}(\epsilon, \epsilon^{\prime}) \bigr\rangle \bigr) \,.
\label{spinCurrent}
\end{equation}

The order parameters are defined from the conventional self-consistency equations

\begin{align}
\Delta _{1,2}&= \frac{\lambda _{S}}{8}\int_{0}^{\theta
_{D}}d\epsilon \,
\mathrm{Tr}\bigl( (\hat{\rho}_0 \pm \hat{\rho}_3) \cdot i\hat{%
\tau}_{2}\cdot \hat{\sigma}_{3} \cdot
\bigl(\check{g}(\epsilon,\epsilon^{%
\prime })\bigr)^{K}\bigr)\,, \label{DeltaSC} \\
W_{Mz}&= \frac{\lambda _{M}}{8}\int_{0}^{\theta _{M}}d\epsilon \,
\mathrm{Tr%
}\bigl(\hat{\rho}_{1}\cdot i\hat{\tau}_{2}\cdot \hat{\sigma}_{3}
\cdot \bigl(%
\check{g}(\epsilon ,\epsilon^{\prime })\bigr)^{K}\bigr)\,.
\label{mSC}
\end{align}

Eqs.~(\ref{Eilenberger}--\ref{DefnLambda}), together with Eqs.~(\ref{DeltaSC}--\ref{mSC})
and the expressions for the electric current, Eq.~(\ref{Current}), and magnetization,
Eqs.~(\ref{Mx,y}--\ref{Mz}), allow one to study various problems both in homogeneous
and nonhomogeneous systems.

In the next section we calculate the magnetic moment induced in the system
by an applied magnetic field and find the Knight shift.

\section{The Knight shift}

In the experiments, \cite{Knight} the nuclear magnetic resonance (NMR) in
superconducting (Sn) granules was studied. It was found that at low
temperatures $T$ the resonance line is shifted with respect to its position
in the absence of the electron polarization (the so called Knight shift).
Since at low $T$ the free electrons in tin are bound in singlet Cooper pairs,
they cannot contribute to the magnetic moment of the granules. Abrikosov and
Gor'kov \cite{AbrikGor} suggested an explanation for the observed Knight shift
taking into account the spin-orbit interaction. They showed that even at zero
temperature this interaction gives rise to a non-zero polarization of electron
spins in an external magnetic field $\mathbf{H}$.

In this section we study the Knight shift in a superconductor with an SDW and
show that even in the absence of the spin-orbit interaction the Knight shift
is finite provided the magnetic field $\mathbf{H}$ is not parallel to the
direction of the magnetization in the SDW.

In order to calculate the electron spin polarization in the field $\mathbf{H}$
we use Eq.~(\ref{Eilenberger}) for the retarded (or advanced Green's functions)
written in the Matsubara representation, Eq.~(\ref{EilenbergerMatsubara}). Since
we consider the uniform case, the second term in Eq.~(\ref{EilenbergerMatsubara})
may be omitted. Thus, we have to solve the equation

\begin{equation}
\bigl\lbrack \check{g} \,, \hat{\mathit{\Pi}}^{(j)}_{\pm} \bigr\rbrack = \bigl\lbrack
\check{g} \,, i W_{Z} \hat{\mathit{P}} \bigr\rbrack \,,
\label{EilenKnight}
\end{equation}%
where

\begin{widetext}
\begin{equation}
\hat{\mathit{\Pi}}^{(j)}_{\pm} = \omega_{n} \hat{\tau}_{3} + i \delta
    \mu \cdot \hat{\rho}_{3} \cdot \hat{\tau}_{0} \cdot \hat{\sigma}_{0} +
    \Delta^{\prime} \cdot \hat{\rho}_{0,3} \cdot \hat{\tau}_{2} \cdot \hat{\sigma}_{3} +
    \Delta^{\prime \prime} \cdot \hat{\rho}_{3,0} \cdot \hat{\tau}_{1} \cdot \hat{\sigma}_{3} +
    W_{M0} \cdot
    \begin{cases}
        - \hat{\rho}_{2} \cdot \hat{\tau}_{1} \cdot \hat{\sigma}_{1} \,, & j=x \\
        \hat{\rho}_{2} \cdot \hat{\tau}_{1} \cdot \hat{\sigma}_{2} \,, & j=y \\
        - \hat{\rho}_{1} \cdot \hat{\tau}_{2} \cdot \hat{\sigma}_{3} \,, & j=z \\
    \end{cases}
\label{PiKnight}
\end{equation}
\end{widetext}
and $\omega _{n}=\pi T(2n+1)$ is the Matsubara frequency. The matrix $%
\mathit{\hat{P}}$ is defined in Eq.~(\ref{DefnLambda}).

For simplicity, we neglect the deviation from the perfect nesting and set
$\delta \mu=0 $. We also consider the case of the $s_{+-}$-pairing and the
magnetization being oriented along the $z$-axis ($j=z$). The energy of
the Zeeman splitting is assumed to be small, $W_{Z}~\ll~\{\Delta,W_{M0}\}$,
which allows us to consider the right-hand side of Eq.~(\ref{EilenKnight})
as a perturbation.

\begin{figure*}[t]
\begin{center}
\includegraphics[width=0.9\textwidth]{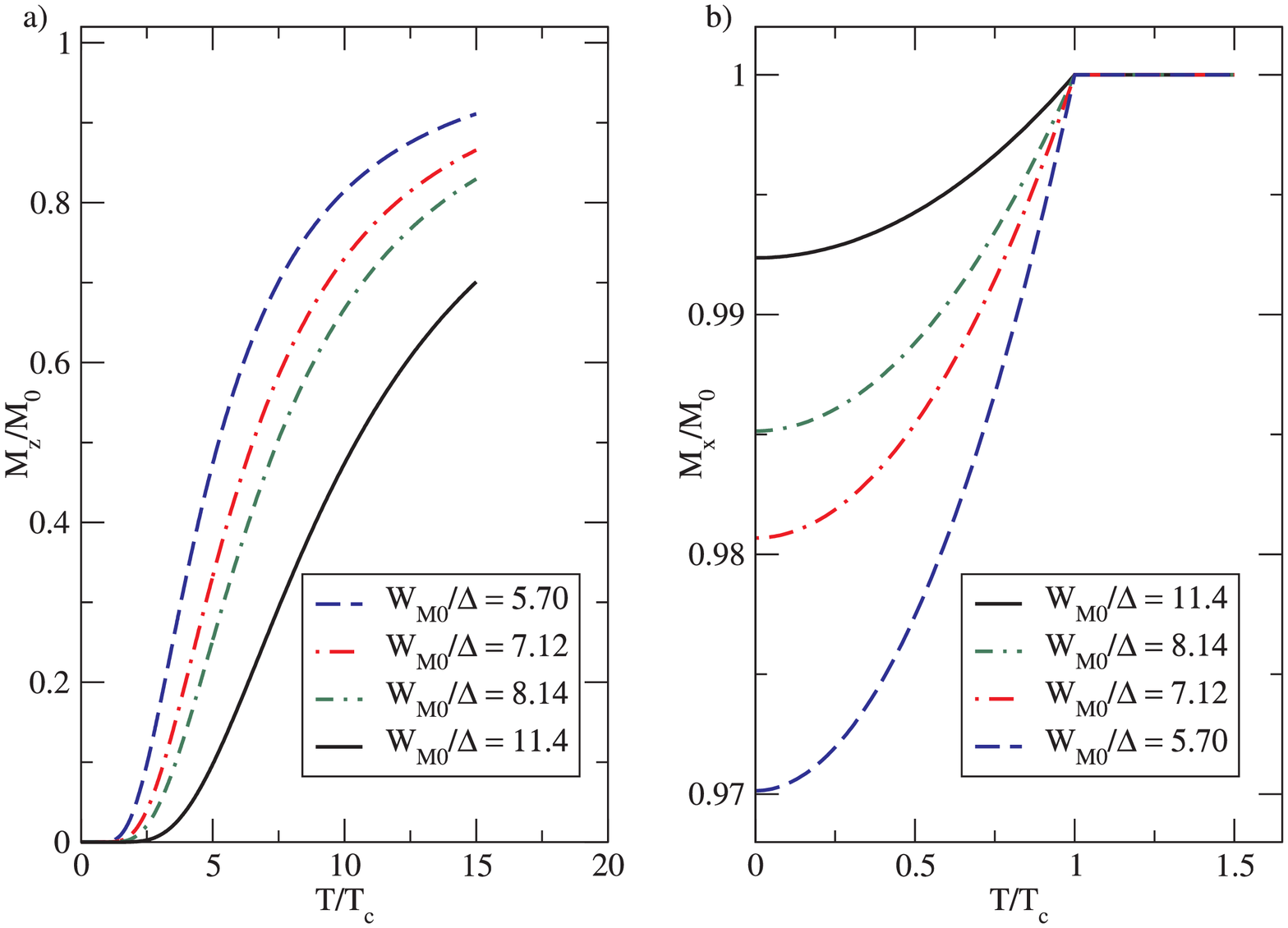}
\end{center}
\caption{(Color online.) Temperature dependence of the magnetic moment induced by a weak magnetic field for different ratios of magnetic to superconducting order parameters.\\
a) $M_{z}(T)$: $M_{z} \rightarrow 0$ as $T \rightarrow 0$;\\
b) $M_{x}(T)$: $M_{x}$ stays finite as $T \rightarrow 0$.}
\label{fig:MzMx}
\end{figure*}

In the zeroth order in $W_{Z}$ we neglect the R.H.S. of Eq.~(\ref{EilenKnight})
and obtain the homogeneous equation

\begin{equation}
\bigl\lbrack \check{g}_{0} \,, \hat{\mathit{\Pi}}^{(z)}_{-} \bigr\rbrack = 0 \,.
\label{EilenKnight0}
\end{equation}

In principle, any matrix function of $\hat{\mathit{\Pi}}_{-}^{(z)}$
satisfies Eq.~(\ref{EilenKnight0}). In order to find the function
$\check{g}_{0}$ unambiguously one should check whether the solution satisfies
the normalization imposed by Eq.~(\ref{NormCond}) or not. This leads us to the solution

\begin{equation}
\check{g}_{0}=\frac{1}{\mathcal{E}_{S}} \hat{\mathit{\Pi}}_{-}^{(z)} \,,
\label{Knightg0}
\end{equation}%
where $\mathcal{E}_{S}^{2}=\omega _{n}^{2}+W_{M0}^{2}+|\Delta |^{2}$.
It is easy to see that the solution given by Eq.~(\ref{Knightg0}) really satisfies
the normalization condition

\begin{equation}
\check{g}_{0} \cdot \check{g}_{0}=\check{1} \,.
\label{KnightNorm0}
\end{equation}

The correction $\delta \check{g} = \check{g} - \check{g}_{0}$ obeys the
equation

\begin{equation}
\bigl\lbrack \delta \check{g} \,, \hat{\mathit{\Pi}}^{(z)}_{-} \bigr\rbrack =
\mathcal{E}_{S} \lbrack \delta \check{g} \,, \check{g}_{0} \rbrack = -2
\mathcal{E}_{S} \check{g}_{0} \cdot \delta \check{g} = i W_{Z} \bigl\lbrack
\check{g}_{0} \,, \hat{\mathit{P}} \bigr\rbrack \,.
\label{KnightCorr}
\end{equation}
Here, we used the relation

\begin{equation}
\delta \check{g} \cdot \check{g}_{0} + \check{g}_{0} \cdot \delta \check{g}
= 0 \,,
\label{KnightNorm}
\end{equation}
which follows from the normalization condition.

Using Eq.~(\ref{KnightNorm0}) we find from Eq.~(\ref{KnightCorr})

\begin{equation}
\delta \check{g} = -\frac{i W_{Z}}{2 \mathcal{E}_{S}} \bigl(\check{g}_{0} \cdot
\hat{\mathit{P}} \cdot \check{g}_{0} - \hat{\mathit{P}}\bigr) \,.
\label{KnightDelG}
\end{equation}

The matrix $\delta \check{g}$ can be written in an explicit form with the
help of Eq.~(\ref{Knightg0}) and the expression for $\hat{\mathit{P}}$. We
present here those parts of $\delta \check{g}$ that contribute to the
magnetic moments, i.e.\ $\delta \check{g}_{x}$ and $\delta \check{g}_{z}$:

\begin{align}
\delta \check{g}_{x} \equiv \delta \check{g}_{1} &= -i W_{Z}
    \frac{|\Delta|^{2}}{\mathcal{E}_{S}^{3}} \hat{\rho}_{3} \cdot \hat{\tau}_{0} \cdot
    \hat{\sigma}_{1} \sin\theta \,,
    \label{KnightCorrx} \\
\delta \check{g}_{z} \equiv \delta \check{g}_{3} &= -i W_{Z}
    \frac{|\Delta|^{2} + W_{M0}^{2}}{\mathcal{E}_{S}^{3}} \hat{\rho}_{0} \cdot
    \hat{\tau}_{3} \cdot \hat{\sigma}_{3} \cos\theta \,.
    \label{KnightCorrz}
\end{align}

Using Eqs.~(\ref{Mx,y}--\ref{Mz}) we find for the spin polarization induced
by the field $\mathbf{H}_{ex}$

\begin{align}
M_{x} &= M_{0} \left(1 - 2 \pi T \sum_{\omega \geqslant 0}
    \frac{|\Delta|^{2}}{\mathcal{E}_{S}^{3}}\right) \sin\theta \,,
\label{ShiftTx} \\
M_{z} &= M_{0} \left(1 - 2 \pi T \sum_{\omega \geqslant 0}
    \frac{|\Delta|^{2} + W_{M0}^{2}}{\mathcal{E}_{S}^{3}}\right) \cos\theta \,.
\label{ShiftTz}
\end{align}

In principle, the sums over $\omega$ in Eqs.~(\ref{ShiftTx}, \ref{ShiftTz})
can be calculated at arbitrary temperatures but the final expressions are
somewhat cumbersome. Therefore, we restrict ourselves by the limit of low
temperatures $T~\rightarrow~0$. In this limit one replaces the sums over
$\omega $ by integrals, which leads to the following expressions

\begin{align}
M_{x} &= M_{0}\frac{W_{M0}^{2}}{|\Delta |^{2}+W_{M0}^{2}}\sin \theta \,,
\label{Shift0x} \\
M_{z} &= M_{0}\left( 1-\int_{0}^{\infty }\frac{W_{M0}^{2}+|\Delta
    |^{2}}{\mathcal{E}_{S}^{3}}d\omega \right) \cos \theta \rightarrow 0\,.
\label{Shift0z}
\end{align}

Eq.~(\ref{Shift0z}) shows that in the limit of the perfect nesting ($\delta\mu~=~0$)
the Knight shift vanishes at $T~=~0$ provided the applied magnetic field $\mathbf{H}$
is parallel to the orientation of the magnetization of the SDW. If the direction of
$\mathbf{H}$ deviates from the $z$-direction, the Knight shift is finite and the
induced magnetic moment $M_{x}$ approaches the spin magnetic moment of free electrons
(Pauli paramagnetism) for $W_{M0}~\gg~\Delta $. The obtained results do not depend on
the relation between $\Delta_{1}$ and $\Delta_{2}$ and thus are valid for both $s_{+-}$-
and $s_{++}$- pairing.

In Fig.(\ref{fig:MzMx}) we plot the temperature dependence of $M_{x}$ and $M_{z}$ for
different ratios $W_{M0}/\Delta$ assuming that in the considered temperature range the
magnetization $\mathbf{m}$ of the SDW depends only weakly on $T$.

At present, it is not easy to quantitatively compare our results obtained within a
simplified model with available experimental data concerning the NMR studies in
pnictides \cite{Chu,Chu1}. There are several reasons for the difficulty and the major
one is that the Knight shift is actually not discussed in those papers. Furthermore,
the influence of the free electrons on the position of the NMR peak in the compound
NaFeAs containing $^{23}$Na atoms is weak because free electrons move in the FeAs
planes. On the other hand, an internal local magnetic field of the SDW causes a stronger
influence on the NMR peaks corresponding to the atoms $^{75}$As. In addition, the ideal
nesting is assumed in our model that leads to the fully gapped Fermi surface. This
assumption is not fulfilled in materials studied in Refs.~\cite{Chu,Chu1}.

\section{Proximity Effect}

We study the proximity effect considering a simple model: a contact between
a superconductor with SDW (two order parameters: $\Delta $ and $W_{M}$),
which we denote as $S_{SDW}$, and a conductor (or insulator at $W_{M}~\neq~0$
and low temperatures) with one order parameter (for example $W_{M}$) or with
a simple normal metal $N$ ($\Delta~=~W_{M}~=~0$). Such a case may be realized
in pnictides with a nonuniform doping level. We will find the quasiclassical
retarded and advanced Green's functions describing the equilibrium properties
such as the density-of-states or the order parameters, $\Delta $ and $W_{M}$.

These Green's functions obey the generalized Eilenberger equation. This
equation is obtained from Eq.~(\ref{Eilenberger}) by taking its element $(11)$
or $(22)$ and performing the Fourier transformation in the Matsubara
representation.

As a result, we obtain

\begin{equation}
n_{x} v \partial_{x} \check{g} - \bigl\lbrack \check{g}\,, \hat{\mathit{\Pi}}_{\pm}^{(j)} \bigr\rbrack = 0\,,
\label{Eilenberger1}
\end{equation}
where $n_{x}=v_{x}/v=\cos\alpha$, $\alpha$ is the angle between the Fermi velocity
$\mathbf{v}$ and the $x$-axis, and $v~=~|\mathbf{v}|$. The matrix
$\hat{\mathit{\Pi}}_{\pm}^{(j)}$ is specified in Eq.~(\ref{PiKnight}).

We assume the $s_{+-}$-pairing and let the magnetization be directed along the
$z$-axis. For simplicity, we assume as previously perfect nesting by putting
$\delta\mu~=~0$.

Two different cases will be considered now:

\begin{itemize}
\item[a)]   $\Delta^{\prime}(x) = \Delta_{0}$, $\Delta^{\prime \prime}(x) = 0$,
            $W_{M0}(x) = W_{M0,S}$ for $x<0$ and
            $\Delta^{\prime}(x) = \Delta^{\prime \prime}(x) = 0$, $W_{M0}(x)=W_{M0,M}$ for $x>0$
\label{Case a}
\item[b)]   $\Delta^{\prime}(x) = \Delta_{0}$, $\Delta^{\prime \prime}(x) = 0$,
            $W_{M0}(x) = W_{M0,S}$ for $x<0$ and
            $\Delta^{\prime}(x) = \Delta^{\prime \prime}(x) = 0$, $W_{M0}(x)=0$ for $x>0$
\label{Case b}
\end{itemize}

The second case corresponds to an interface between a superconductor with
SDW and a normal (nonmagnetic) metal. We denote this type of contacts as $S_{SDW}/N$.

The first case corresponds to a system with SDW having the superconducting
order parameter at $x<0$. This type of contact is denoted as $S_{SDW}/N_{SDW}$.

The contact between the two regions is assumed to be ideal and therefore all
the functions should be continuous across the boundary $x~=~0$.

As in the previous section, we represent the solution $\check{g}(x)$ in the form

\begin{equation}
\check{g}(x)=\check{g}_{0}+\delta \check{g}(x)\,.
\label{g(x)}
\end{equation}
Here the matrix $\check{g}_{0}$ is a constant in space and obeys the equation

\begin{equation}
\bigl\lbrack \check{g}_{0} \,, \hat{\mathit{\Pi}}^{(z)}_{-} \bigr\rbrack = 0 \,.
\label{Eqg0}
\end{equation}
The proper solution of this equation is written again in the form

\begin{equation}
\check{g}_{0} = \frac{1}{\mathcal{E}_{S}}\hat{\mathit{\Pi}}^{(z)}_{-}
\label{g0}
\end{equation}
with $\mathcal{E}_{S}^{2} = \omega_{n}^{2} + \Delta_{0}^{2} + W_{M0}^{2}$.

The matrix $\delta \check{g}(x)$ is not supposed to be small. It can be
split into an even, $\check{s}$, and odd, $\check{a}$, in $n_{x}$ parts

\begin{equation}
\delta \check{g} = \check{s} + n_{x} \check{a} \,.
\label{s+a}
\end{equation}

Substituting Eq.~(\ref{s+a}) into Eq.~(\ref{Eilenberger1}) and separating the
even and odd in $n_{x}$ parts, we come to equations

\begin{align}
v \partial_{x} \check{s} + \bigl\lbrack \hat{\mathit{\Pi}}^{(z)}_{-} \,,
    \check{a} \bigr\rbrack &= 0 \,,
\label{Eqs} \\
n_{x}^{2} v \partial_{x} \check{a} + \bigl\lbrack \hat{\mathit{\Pi}}^{(z)}_{-} \,,
    \check{s} \bigr\rbrack &= 0 \,.
\label{Eqa}
\end{align}

One can exclude the anisotropic part $\check{a}$ by differentiating Eq.~(\ref{Eqs})
with respect to the coordinate $x$. Then, the equation for the isotropic part has the form

\begin{equation}
-l_{\alpha S}^{2} \frac{\partial^{2} \check{s}}{\partial x^{2}} + \frac{1}{2\mathcal{E}_{S}^{2}}
\Bigl(\mathcal{E}_{S}^{2}\check{s} -
\hat{\mathit{\Pi}}^{(z)}_{-} \check{s} \hat{\mathit{\Pi}}^{(z)}_{-}\Bigr) = 0 \,,
\label{Eq-s}
\end{equation}
where $l_{\alpha S}=n_{x}v/2\mathcal{E}_{S}$ is a characteristic length of
penetration of perturbations caused by the proximity effect into the
superconductor with SDW. At low temperatures this length is determined by
the smallest of the lengths $\{\xi_{\Delta}\simeq v/\Delta_{0}, \xi_{M}\simeq v/W_{M0,M}\}$.

In the region $x>0$ the characteristic length is $l_{\alpha M}~=~n_{x}v/2\mathcal{E}_{M}$
(the case a)) or $l_{\alpha N}~=~n_{x}v/2\mathcal{E}_{\omega}$ (the case b)), where
$\mathcal{E}_{M}^{2}~=~\omega_{n}^{2}~+~W_{M0,M}^{2}$ and $\mathcal{E}_{\omega}^{2}~=~\omega_{n}^{2}$.

We look for a solution of Eq.~(\ref{Eq-s}) in the form

\begin{equation}
\check{s}=\begin{cases}
        A(x)\hat{\tau}_{3}+B(x)\hat{\rho}_{3}\cdot \hat{\tau}_{2}\cdot
            \hat{\sigma}_{3}+C(x)\hat{\rho}_{1}\cdot \hat{\tau}_{2}\cdot \hat{\sigma}_{3}\,, & x<0 \\
        \tilde{A}(x)\hat{\tau}_{3}+\tilde{B}(x)\hat{\rho}_{3}\cdot \hat{\tau}_{2}
            \cdot \hat{\sigma}_{3}+\tilde{C}(x)\hat{\rho}_{1}\cdot \hat{\tau}_{2}
            \cdot \hat{\sigma}_{3}\,, & x>0
        \end{cases}
\label{a1}
\end{equation}

Substituting Eq.~(\ref{a1}) into Eq.~(\ref{Eq-s}) one can rather easily find the
functions $A$, $B$, $C$, etc.

\begin{align}
A(x) &= A_{0}\exp (x/l_{\alpha S})\,, &\qquad \tilde{A}(x) &= \tilde{A}_{0}\exp (-x/l_{\alpha M})\,, \\
B(x) &= B_{0}\exp (x/l_{\alpha S})\,, &\qquad \tilde{B}(x) &= \tilde{B}_{0}\exp (-x/l_{\alpha M})\,, \\
C(x) &= C_{0}\exp (x/l_{\alpha S})\,, &\qquad \tilde{C}(x) &= \tilde{C}_{0}\exp (-x/l_{\alpha M})\,,
\end{align}

\begin{align}
A_{0} \omega_{n} + B_{0}\Delta_{0} + C_{0} W_{M0,S} &= 0\,, \\
\tilde{A}_{0}\omega_{n} + \tilde{C}_{0} W_{M0,S} &= 0\,.
\label{A+B+C}
\end{align}

We see that there are four independent arbitrary constants: $B_{0}$, $C_{0}$,
$\tilde{B}_{0}$ and $\tilde{C}_{0}$ characterizing the solution $\check{s}$.
They should be determined from the matching conditions that require the
continuity of the matrices $\check{g}_{0}+\check{s}$ and $\check{a}$. The
matching conditions are reduced to six equations four of which are independent
and the other two follow from these four equations. Solving these equations, we find

\begin{align}
A_{0} &= \frac{\varsigma \varpi_{M} - \varpi_{S}}{1+\varsigma} \,, &\qquad \tilde{A}_{0} &= \frac{\varsigma \varpi_{S}-\varpi_{M}}{1+\varsigma}\,, \\
B_{0} &= -\frac{1}{1+\varsigma} \Delta_{0} \,, &\qquad \tilde{B}_{0} &= \frac{\varsigma}{1+\varsigma} \Delta_{0} \,, \\
C_{0} &= \frac{\varsigma \mathit{m}_{M}-\mathit{m}_{S}}{1+\varsigma} \,, &\qquad \tilde{C}_{0} &= \frac{\varsigma \mathit{m}_{S}-\mathit{m}_{M}}{1+\varsigma} \,.
\label{ABC1}
\end{align}
where $\varsigma ~=~(\varpi_{S}\varpi_{M}~+~\mathit{m}_{S}\mathit{m}_{M})^{-1}$,
$\varpi_{S}~=~\omega /\mathcal{E}_{S}$, $\varpi_{M}~=~\omega /\mathcal{E}_{M}$
are the normalized Matsubara frequencies in the $SC$ and $M$ regions, and
$\mathit{m}_{S,M}~=~W_{M0S,M}/\mathcal{E}_{S,M}$ are the magnetic order parameters
in these regions. In the case of a contact of a superconductor with SDW and of a
normal metal ($S_{SDW}/N$ contact), the energy $\mathcal{E}_{M}$ should be replaced
by $\mathcal{E}_{\omega}$ and the quantity $W_{M0}$ set to be zero.

The amplitudes $\tilde{B}_{0}$, $\tilde{C}_{0}$ determine the penetration of the
superconducting and magnetic correlations into the region with SDW or into the normal metal
$N$ due to the proximity effect. The amplitudes $B_{0}$, $C_{0}$ describe the
inverse proximity effect or, in other words, a suppression of $\Delta_{0}$ and $W_{M0}$
in the superconductor near the $S_{SDW}/N_{SDW}$ interface (or in the $S_{SDW}/N$
interface) due to the inverse proximity effect.

Note that, strictly speaking, in the case of the $S_{SDW}/N_{SDW}$ system we have
to calculate the magnetic order parameter $M_{M0}$ self-consistently using the
amplitude $\tilde{C}_{0}$. This makes the problem more difficult. However, the
obtained results remain valid provided the quantity $W_{M0}$ is the same at $x>0$
and $x<0$ (i.e.\ $W_{M0,S}~=~W_{M0,M}~\equiv~W_{M0}$). In this case we obtain

\begin{align}
A_{0} &= -\frac{\omega \Delta_{0}^{2}}{\mathcal{E}_{S}\mathcal{E}_{M}(\mathcal{E}_{S}+\mathcal{E}_{M})}\,, &\qquad
    \tilde{A}_{0} &= 0\,, \\
B_{0} &= -\frac{\Delta_{0} \mathcal{E}_{M}}{\mathcal{E}_{S}(\mathcal{E}_{S}+\mathcal{E}_{M})}\,, &\qquad
    \tilde{B}_{0} &= \frac{\Delta_{0}}{\mathcal{E}_{S}+ \mathcal{E}_{M}}\,, \\
C_{0} &= \frac{m_{0}\Delta_{0}^{2}}{\mathcal{E}_{S}\mathcal{E}_{M}(\mathcal{E}_{S}+\mathcal{E}_{M})}\,, &\qquad
    \tilde{C}_{0} &= 0\,.
\label{ABC2Tilde}
\end{align}

The results obtained mean that the corrections to the DOS and to the magnetic order
parameter determined by $\tilde{A}_{0}$ and $\tilde{C}_{0}$ are absent in this
case. The superconducting pair function $\tilde{B}(x)$ penetrates the region
with SDW over the length $l_{\alpha M}$ with the amplitude $\tilde{B}_{0}$. As
it should be, this pair function penetrates the $N$ region over the length
$l_{\alpha N}~=~l_{\alpha \omega}~=~|n_{x}|v/\omega_{n}$.

\section{Josephson Effect}

In this section, we calculate the dc Josephson current in an $S_{SDW}/N/S_{SDW}$
system using a simple model. We assume again an
ideal nesting ($\delta \mu =0$) and take into account the impurity scattering in the
self-consistent Born approximation. Then, the Eilenberger equation, Eq.~(\ref{Eilenberger1}),
for the matrix $\check{g}$ acquires the form
\begin{equation}
n_{x}v\partial_{x} \check{g} - \bigl\lbrack \check{g} \,, \hat{\mathit{\Pi}} \bigr\rbrack =
    \frac{1}{2\tau} \bigl\lbrack \check{g} \,, \langle \check{g} \rangle \bigr\rbrack \,,
\label{Jos1}
\end{equation}
where the angle brackets stand for the average over the directions of the momentum and
the matrix $\hat{\mathit{\Pi}}$ in the right (left) superconductors is equal to

\begin{widetext}
\begin{equation}
\hat{\mathit{\Pi}} = \omega_{n} \hat{\tau}_{3} + \Delta \left(\hat{\rho}_{0,3}\cdot
\hat{\tau}_{2} \cdot \hat{\sigma}_{3} \cos\left(\frac{\varphi}{2}\right) \pm
\hat{\rho}_{3,0} \cdot \hat{\tau}_{1} \cdot \hat{\sigma}_{3}
\sin\left(\frac{\varphi}{2}\right) \right) +
W_{M0} \left( \hat{\rho}_{1} \cdot \hat{\tau}_{2} \cdot \hat{\sigma}_{3}
\cos\left(\frac{\alpha}{2}\right) \mp
\hat{\rho}_{2} \cdot \hat{\tau}_{1} \cdot \hat{\sigma}_{2}
\sin\left(\frac{\alpha}{2}\right) \right) \,.
\label{Jos2}
\end{equation}
\end{widetext}

\begin{figure}[t]
\begin{center}
\includegraphics[width=0.45\textwidth]{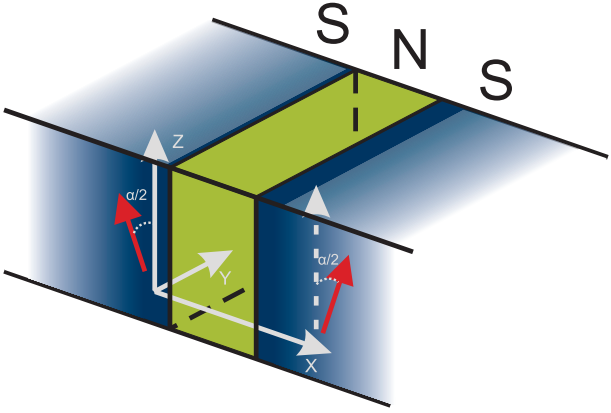}
\end{center}
\caption{(Color online.) Josephson junction under consideration. A normal metal film is placed between two
superconducting leads with SDW. The orientation of the magnetization vectors of
the SDW in the leads lying in the $y$-$z$-plane is shown in red.}
\label{fig:JosephsonSystem}
\end{figure}

Eq.~(\ref{Jos2}) corresponds to the case when the superconducting phase in the
right (left) $S_{SDW}$ equals $\pm \varphi/2$ and the angle between the vector
of the magnetization of the SDW and $z$-axis is equal to $\pm\alpha/2$
(cf.\ Fig.~(\ref{fig:JosephsonSystem})).
Considering the impurity scattering, we neglect interband scattering and regard,
for simplicity, the impurity scattering time $\tau$ equal for each band. In the
middle of the $N$ layer $\Delta~=~0$ and $W_{M0}~=~0$. We assume that the scattering
time $\tau$ in this layer is rather short: $\hbar \tau^{-1}~\gg~T$ (diffusive limit).

As a boundary condition, we adopt the one obtained for a simplified model suggested
in Ref.~\cite{KL}

\begin{equation}
\check{g}\left. \frac{\partial \check{g}}{\partial x}\right| _{x=\pm
L}=\pm
\gamma
_{B}\bigl\lbrack\check{g}\,,\check{g}_{S}\bigr\rbrack\biggr|_{x=\pm
L}\,, \label{Jos3}
\end{equation}
where the parameter $\gamma_{B}~=~(2 R_{B}\sigma_{N})^{-1}$ depends on the interface
resistance $R_{B}$ per unit area and the conductivity $\sigma_{N}$ of the normal metal.
This parameter is assumed to be a scalar, which means that we neglect transitions
between different bands at the interfaces. In a more general case $\gamma_{B}$ is a
matrix \cite{Millis,EschrigBC,NazarovBC}.

We assume that the proximity effect is weak, which corresponds to small values of the
parameter $\gamma_{B}\xi_{N}$, where $\xi _{N}~=~\sqrt{D/2\pi T}$. In this case the
Green's functions in the $S_{SDW}$ are only weakly perturbed by the contact with the
$N$ layer (because of a large interface resistance) and the first term in Eq.~(\ref{Jos1})
can be neglected. Then, the solution of Eq.~(\ref{Jos1}) on the boundaries between
the normal metal and superconductors can be written as

\begin{equation}
\check{g}_{S}(\pm L)=\check{g}_{+} \pm \check{g}_{-} \,,
\label{Jos4}
\end{equation}
where

\begin{align}
\check{g}_{+} = \frac{1}{\mathcal{E}_{S}} \biggl\lbrack \omega_{n}\hat{\tau}_{3} &+
    \Delta \hat{\rho}_{0,3} \cdot \hat{\tau}_{2} \cdot \hat{\sigma}_{3}
    \cos\left(\frac{\varphi}{2}\right)
\label{Jos5} \\
    &+ W_{M0} \hat{\rho}_{1} \cdot \hat{\tau}_{2} \cdot \hat{\sigma}_{3}
    \cos\left(\frac{\alpha }{2}\right) \biggr\rbrack \notag
\end{align}
and

\begin{align}
\check{g}_{-}=\frac{1}{\mathcal{E}_{S}}\biggl\lbrack&  \Delta
\hat{\rho}%
_{3,0}\cdot \hat{\tau}_{1}\cdot \hat{\sigma}_{3}\sin \left(
\frac{\varphi }{2%
}\right) \label{Jos6} \\
&  -W_{M0}\hat{\rho}_{2}\cdot \hat{\tau}_{1}\cdot
\hat{\sigma}_{2}\sin \left(
\frac{\alpha }{2}\right) \biggr\rbrack \notag
\end{align}
with $\mathcal{E}_{S}^{2}=\omega_{n}^{2}+W_{M0}^{2}+\Delta ^{2}$.

To find the Josephson current $\mathbf{j}_{J}$, we have to solve
Eq.~(\ref{Jos1}) in the $N$ layer, where $\Delta~=~0$ and $W_{M0}~=~0$. As
usually, we represent the Green's function $\check{g}$ as a sum of the
symmetric and antisymmetric functions: $\check{g}~=~\check{s}~+~n_{x}\check{a}$
(see Eq.~(\ref{s+a})). For $\check{s}$ and $\check{a}$ we obtain from Eq.~(\ref{Jos1})
the following equations

\begin{align}
v \partial_{x} \check{s} + \omega_{n} \lbrack \hat{\tau}_{3} \,, \check{a} \rbrack &=
    -\frac{1}{2\tau} \left( \check{s}\check{a} - \check{a}\check{s} \right) \,,
\label{Jos7} \\
v \langle n_{x}^{2} \partial_{x} \check{a} \rangle + \omega_{n} \bigl\lbrack \hat{\tau}_{3} \,,
    \langle \check{s} \rangle \bigr\rbrack &= 0 \,,
\label{Jos8}
\end{align}
where the angle brackets denote the angle averaging. In the diffusive limit
considered here ($\tau T/\hbar~\ll~1$) the second term in Eq.~(\ref{Jos7}) in
the left hand side is small. The right-hand side can be transformed using
the equation

\begin{equation}
\check{s}\check{a}+\check{a}\check{s}=0 \,,
\label{Jos9}
\end{equation}
which follows immediately from the normalization condition Eq.~(\ref{NormCond}).
Then, we obtain for $\check{a}$

\begin{equation}
\check{a}=-l\check{s} \, \partial_{x} \check{s} \,.
\label{Jos10}
\end{equation}
We used another part of the normalization condition (for symmetric function)

\begin{equation}
\check{s}^{2} + n_{x}^{2} \check{a}^{2} = 1 \,,
\label{Jos11}
\end{equation}
in which the second term can be neglected (as follows from Eq.~(\ref{Jos11})
$|\check{a}|~\ll~1)$. In the considered limit of a weak proximity effect
($|\check{s}|~\ll~1$) the matrix $\check{s}$ can be represented in the form

\begin{equation}
\check{s} = \hat{\tau}_{3} \, \mathrm{sgn}(\omega) + \delta \check{s} \,,
\label{Jos12}
\end{equation}
where the matrix $\delta\check{s}\approx\langle\delta\check{s}\rangle$
obeys the equation

\begin{equation}
- \frac{\partial^{2} \delta \check{s}}{\partial x^{2}} + \kappa_{\omega}^{2} \delta \check{s} = 0
\label{Jos13}
\end{equation}
with $\kappa_{\omega}^{2}=2\omega_{n}/D$, $D=vl/3$ being the diffusion
coefficient which is assumed to be the same in each band. Eq.~(\ref{Jos13})
follows from Eq.~(\ref{Jos8}) because, as we will see, the matrix $\delta \check{s}$
anticommutes with the matrix $\hat{\tau}_{3}$.

Eq.~(\ref{Jos13}) has to be solved with the boundary conditions which follow
from Eq.~(\ref{Jos3}) and the representation Eq.~(\ref{Jos12})

\begin{equation}
\frac{\partial \delta \check{s}}{\partial x}(\pm L) = \pm \gamma_{B}
\left\lbrack \check{g}_{S}(\pm L) - \hat{\tau}_{3} \frac{\omega}{\mathcal{E}_{S}} \right\rbrack \,.
\label{Jos14}
\end{equation}
The solution of Eq.~(\ref{Jos14}) can easily be found in the form

\begin{equation}
\delta \check{s}(x) = \check{A} \cdot \frac{\cosh(\kappa_{\omega} x)}{\sinh(\kappa_{\omega} L)} +
    \check{B} \cdot \frac{\sinh(\kappa_{\omega} x)}{\cosh(\kappa_{\omega} L)} \,,
\label{Jos15}
\end{equation}
where

\begin{align}
\check{A} = \frac{\gamma_{B}}{\mathcal{E}_{S}} &\left\lbrack \Delta \hat{\rho}_{0,3} \cdot
    \hat{\tau}_{2} \cdot \hat{\sigma}_{3} \cos\left(\frac{\varphi}{2}\right) \right.
    \label{Jos16} \\
    + &\left. W_{M0} \hat{\rho}_{1} \cdot \hat{\tau}_{2} \cdot \hat{\sigma}_{3}
    \cos\left(\frac{\alpha}{2})\right) \right\rbrack \notag \\
\intertext{and}
\check{B} = \frac{\gamma_{B}}{\mathcal{E}_{S}} &\left\lbrack \Delta \hat{\rho}_{3,0} \cdot
    \hat{\tau}_{1} \cdot \hat{\sigma}_{3} \sin\left(\frac{\varphi}{2}\right) \right.
    \label{Jos17} \\
    - &\left. W_{M0} \hat{\rho}_{2} \cdot \hat{\tau}_{2} \cdot \hat{\sigma}_{2}
    \sin\left(\frac{\alpha }{2}\right) \right\rbrack \,. \notag
\end{align}

The Josephson current $j_{J}$ can be calculated using Eq.~(\ref{Current}) and the
expression for $\check{a}$ (\ref{Jos10}). We are interested in the part of $\check{a}$
which contributes to the current. This part can be written in the main approximation as

\begin{equation}
\check{a} = -l \, \delta \check{s} \, \partial_{x} \delta \check{s} \,.
\label{Jos18}
\end{equation}

Proceeding in this way we reduce the expression for the Josephson current $j_{J}$ to the form

\begin{equation}
j_{J}=\frac{i \sigma (2\pi T)}{8e}\sum_{\omega} \mathrm{Tr} \bigl\lbrace \hat{\rho}_{3} \cdot
\hat{\tau}_{3} \cdot \hat{\sigma}_{0} \cdot \delta \check{s} \, \partial_{x} \delta \check{s} \bigr\rbrace \,.
\label{Jos19}
\end{equation}
Substituting Eqs.~(\ref{Jos15}--\ref{Jos17}) into Eq.~(\ref{Jos19}), we find

\begin{equation}
j_{J} = j_{c} \sin\varphi \,,
\label{Jos20}
\end{equation}
where the critical Josephson current $j_{c}$ is given by

\begin{equation}
j_{c} = \sigma \gamma_{B}^{2} L (2\pi T) \sum_{\omega} \frac{\Delta^{2}}{\mathcal{E}_{S}^{2}
\theta_{\omega} \sinh (2\theta_{\omega})} \,,
\label{Jos21}
\end{equation}%
where $\theta_{\omega} = \kappa_{\omega} L$. This formula differs from the expression for $j_{c}$
in an $S/N/S$ junction only by the term $W_{M0}^{2}$ in the energy $\mathcal{E}_{S}$. The
presence of this term leads to a suppression of the current $j_{c}$ provided $W_{M0}$ is not small
compared to the superconducting energy gap $\Delta$. Note that, in our simple model, the Josephson
critical current does not depend on the mutual orientation of the vectors of the magnetization in
the left and right superconductors $S_{SDW}$.

It is of interest to calculate the spin current in the Josephson junction. The spin current is
given by Eq.~(\ref{spinCurrent}). Using Eq.~(\ref{Jos10}) one can write the spin current between
the superconductors as

\begin{equation}
j_{sp, x}^{(x)} = -\frac{(2\pi T) D \nu \mu_{B}}{8e} \sum_{\omega} \mathrm{Tr} \bigl\lbrace
\hat{\rho}_{3} \cdot \hat{\tau}_{3} \cdot \hat{\sigma}_{1} \cdot \delta \check{s} \,
\partial_{x} \delta \check{s} \bigr\rbrace \,,
\label{Jos22}
\end{equation}%
where the upper index means the spin orientation, and the lower stands for the direction of the
current. Substituting Eqs.~(\ref{Jos15}--\ref{Jos17}), we obtain from Eq.~(\ref{Jos22})

\begin{equation}
j_{sp, x}^{(x)} = \gamma_{B}^{2} L(2 \pi T) D \nu \mu_{B} \sum_{\omega} \frac{W_{M0}^{2}}{\mathcal{E}_{S}^{2}
\theta_{\omega}\sinh(2 \theta_{\omega})} \sin\alpha \,.
\label{Jos23}
\end{equation}

Eq.~(\ref{Jos23}) resembles the Josephson expression for the supercurrent. Both formulas contain
sine of an angle. In the conventional Josephson formula this angle is equal to the difference of
the phases of the superconductors, whereas the angle $\alpha$ in the expression for the spin
current in Eq.~(\ref{Jos23}) determines the mutual orientation of the SDW in the right and left
electrodes. Eq.~(\ref{Jos22}) is valid also above the superconducting transition temperature
$T_{c}$, when $\Delta~=~0$. This current is dissipationless like the Josephson current. The nature of
a similar dissipationless spin current in systems which differ from ours has been discussed in
Ref.~\cite{MacDonald}

\section{Discussion}

We have derived equations for the quasiclassical Green's functions for a two-band superconductor
with an SDW. It was assumed that the Fermi velocities in the electron and hole bands are equal. We
neglected the anisotropy of the Fermi surfaces.

Using these equations and assuming the ideal nesting, we considered three problems: the Knight shift, the proximity effect and
the dc Josephson effect. It was shown that, provided the direction of the applied magnetic field
coincides with the direction of the magnetization $\mathbf{m}$ in the SDW, the Knight shift
vanishes at zero temperature and in the absence of the spin-orbit interaction. If the magnetic
field is not collinear with the $\mathbf{m}$ vector, the Knight shift is finite, which correlates
with the results of a recent paper, \cite{Ghaemi} where it was shown that the DOS of a superconductor
with the SDW also depends on the orientation of the external magnetic field.

We have demonstrated that near the interface between the superconductor with the SDW and a normal
metal the components of $\check{g}$ describing both magnetic and superconducting correlations
penetrate into the normal metal over the length of the order $\min \{\hbar v/W_{M0}, \hbar v/\Delta \}$.

Using the simplest model of the $S_{SDW}/N/S_{SDW}$ Josephson junction, we calculated the critical
Josephson current and showed that, in this model, it does not depend on the mutual orientation of
the magnetization in the superconductors with the SDW.

However, the dissipationless spin current ($j_{sp}$), which arises in the junction, depends on the misorientation angle
between the magnetisations of the SDW, $\alpha$, in the same way ($j_{sp} \sim \sin \alpha $) as the Josephson current
$j_{J}$ depends on the phase difference of the superconducting order parameter.

Although the equations for the quasiclassical Green's functions have been derived using the simplest
model of a two-band superconductor with an SDW, we believe that the results obtained on the basis of
this model remain valid, at least qualitatively, for more complicated models describing realistic materials.

The derived equations can be easily generalized to the case of impurity scattering and can be applied
to different problems, equilibrium and nonequilibrium (ac Josephson effects in $S_{SDW}/N/S_{SDW}$
junctions, vortices, etc.).

\textit{Note added in proof.} Although Eqs.~(\ref{Eilenberger},
\ref{NormCond})
for the quasiclassical Green's functions for two-band
superconductors with the SDW are valid for arbitrary deviation from
the ideal nesting ($\mu~\neq~0$), we assumed the ideal nesting
($\mu~=~0$) when applying these equations to the study of particular
effects. This assumption is justified, strictly speaking, in a
hypothetical case of equal critical transition temperatures
($T_{cS}~=~T_{cSDW}$) or in the case of a metastable states (when a
first-order transition takes place). In real materials $\mu~\neq~0$,
and therefore one has to take into account the dependence of the
order parameters ($\Delta $ and $W_{M}$) and other quantities on
$\mu $. For example, one can show that the Josephson critical
current $j_{c}$ contains an additional term $\delta j_{c}$ which is
negative and depends on the angle $\alpha$: $\delta j_{c}~\sim~-\mu
^{2}\cos^{2}\alpha $.

\section{Acknowledgements}

The authors are grateful to I. Eremin for useful remarks and discussions. We thank SFB 491 for financial support.

\end{document}